\documentclass[aps,prb,showpacs,reprint,superscriptaddress,floatfix]{revtex4-1}
\usepackage{graphicx}
\begin{document}

\title{\emph{ In-situ\/} measurements of the optical absorption of dioxythiophene-based conjugated polymers}

\author{J. Hwang}

\affiliation{Department of Physics, University of Florida, Gainesville, Florida  32611}

\affiliation{Department of Physics, Pusan National University, Busan 609-735, Republic of 
Korea}

\author{I. Schwendeman}
\altaffiliation{Present address: Department of Physics, University of Texas, Austin, TX}
\affiliation{Department of Chemistry, University of Florida, Gainesville, Florida 32611}

\author{B.C. Ihas}\altaffiliation{Present address: NREL, Boulder, CO} 
\affiliation{Department of Physics, University of Florida, Gainesville, Florida  32611}

\author{R.J. Clark}
\altaffiliation{Present address: Department of Physics, University of Texas, Austin, TX}
\affiliation{Department of Physics, University of Florida, Gainesville, Florida  32611}

\author{M. Cornick}
\altaffiliation{Present address: MIT Lincoln Laboratories Cambridge, MA} 
\affiliation{Department of Physics, University of Florida, Gainesville, Florida  32611}

\author{M. Nikolou}
\altaffiliation{Present address: Cass Business School, City University 
(London), U.K.}
\affiliation{Department of Physics, University of Florida, Gainesville, Florida  32611}

\author{A. Argun}
\altaffiliation{Present address: Giner, Inc., 89 Rumford Avenue, Newton, MA  02466, 
USA}
\affiliation{Department of Chemistry, University of Florida, Gainesville, Florida 32611}

\author{J.R. Reynolds}
\affiliation{Department of Chemistry, University of Florida, Gainesville, Florida 32611}

\author{D.B. Tanner}
\affiliation{Department of Physics, University of Florida, Gainesville, Florida  32611}

\date{\today}

\begin{abstract}

Conjugated polymers can be reversibly doped by electrochemical
means. This doping introduces new sub-bandgap optical absorption
bands in the polymer while decreasing the bandgap absorption. To
study this behavior, we have prepared an electrochemical cell
allowing {\it in situ\/} measurements of the optical properties of
the polymer. The cell consists of a thin polymer film deposited on
gold-coated Mylar behind which is another polymer that serves as a
counterelectrode. An infrared transparent window protects the upper
polymer from ambient air. By adding a gel electrolyte and making
electrical connections to the polymer-on-gold films, one may study
electrochromism in a wide spectral range. As the cell voltage (the
potential difference between the two electrodes) changes, the doping
level of the conjugated polymer films is changed reversibly. Our
experiments address electrochromism in
poly(3,4-ethylene\-dioxy\-thiophene) (PEDOT) and
poly(3,4-dimethyl\-propylene\-dioxy\-thiophene) (PProDOT-Me$_2$). 
This closed electrochemical cell allows the study of the doping
induced sub-bandgap features (polaronic and bipolaronic modes) in
these easily oxidized and highly redox switchable polymers. We also
study the changes in cell spectra as a function of polymer thickness
and investigate strategies to obtain cleaner spectra, minimizing the
contributions of water and gel electrolyte features.

\end{abstract}


\maketitle

\section{Introduction}

Conjugated polymers, especially thiophene-derivative polymers, have
been studied as electrochromic materials for many applications:
smart windows, optical switches, low energy displays,
etc.\cite{pedot1,pedot2,pxdot1,pprodot1,skotheim07,monk07} These
polymers in their doped states have high stability in air and at
high temperatures ($\sim120$~$^{\circ}$C).\cite{pxdot2} Over the
years, several device studies have been done on such electrochromic
polymers.\cite{pedot3,pedot4,pani,pprodot2,argun04,nikolou09}

Electroactive $\pi$-conjugated polymers can be reversibly doped
(oxidized or reduced) to achieve high electrical 
conductivity.\cite{heeger1} Thiophene and its derivatives fall into
the class of non-degenerate ground state
polymers,\cite{theory1,theory2,theory3,theory4,theory5} leading to
polaron or bipolaron states for charged defects in the polymer.
Quantum-chemical calculations of the electronic structure of the
polaron and bipolaron have been done for specific polymers, e.g.,
polyparaphenylene, polypyrrole, and
polythiophene.\cite{theory2,theory3}

\begin{figure}
\includegraphics[width=\hsize]{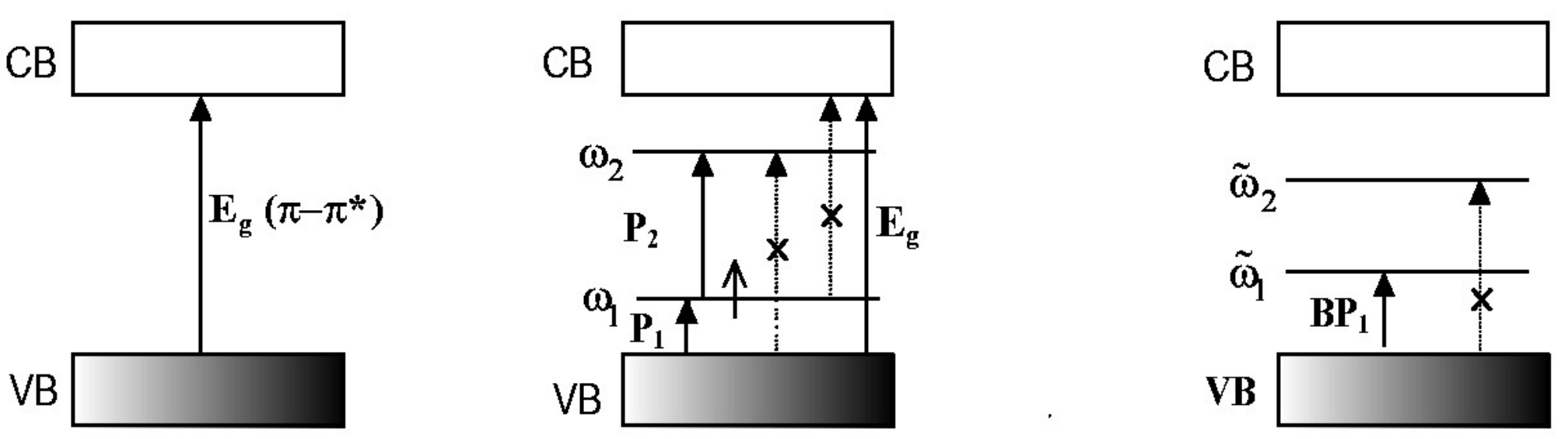}
\caption{\label{ndgsp}
Electronic structure of polarons and bipolarons in p-doped non-degenerate ground state polymers. Vertical lines show the electronic transitions. The lines with $\times$ are for transitions which are not allowed because of symmetry or the
dipole selection rule\cite{dipole}. P$_1$=${\omega}_{1}$, P$_2$=${\omega}_{2}-{\omega}_{1}$, and BP$_1$=${\tilde{\omega}}_{1}$. The small arrow stands for an electron with a spin (either up or down).}
\end{figure}

A key signature of polaron or bipolaron formation in these polymer
systems is the appearance of midgap states and their associated
electronic transitions (polaronic or bipolaronic), which are
revealed by optical absorption
experiments.\cite{kaufman84,Chung84,Harbeke86,pedot3,edot-ito}
Figure~\ref{ndgsp} shows the general features of the doping-induced
electronic structure; the polaron is the expected state when the
system is lightly doped whereas the bipolaron is the expected state
when the system is heavily doped. Polarons and bipolarons exhibit
different absorption spectra: the polaron state yields three broad
peaks whereas the bipolaron state yields only a single, even
broader, peak.

\begin{figure}
\centering
 \vspace*{-0.5 cm}%
 \centerline{\includegraphics[width=3.6 in]{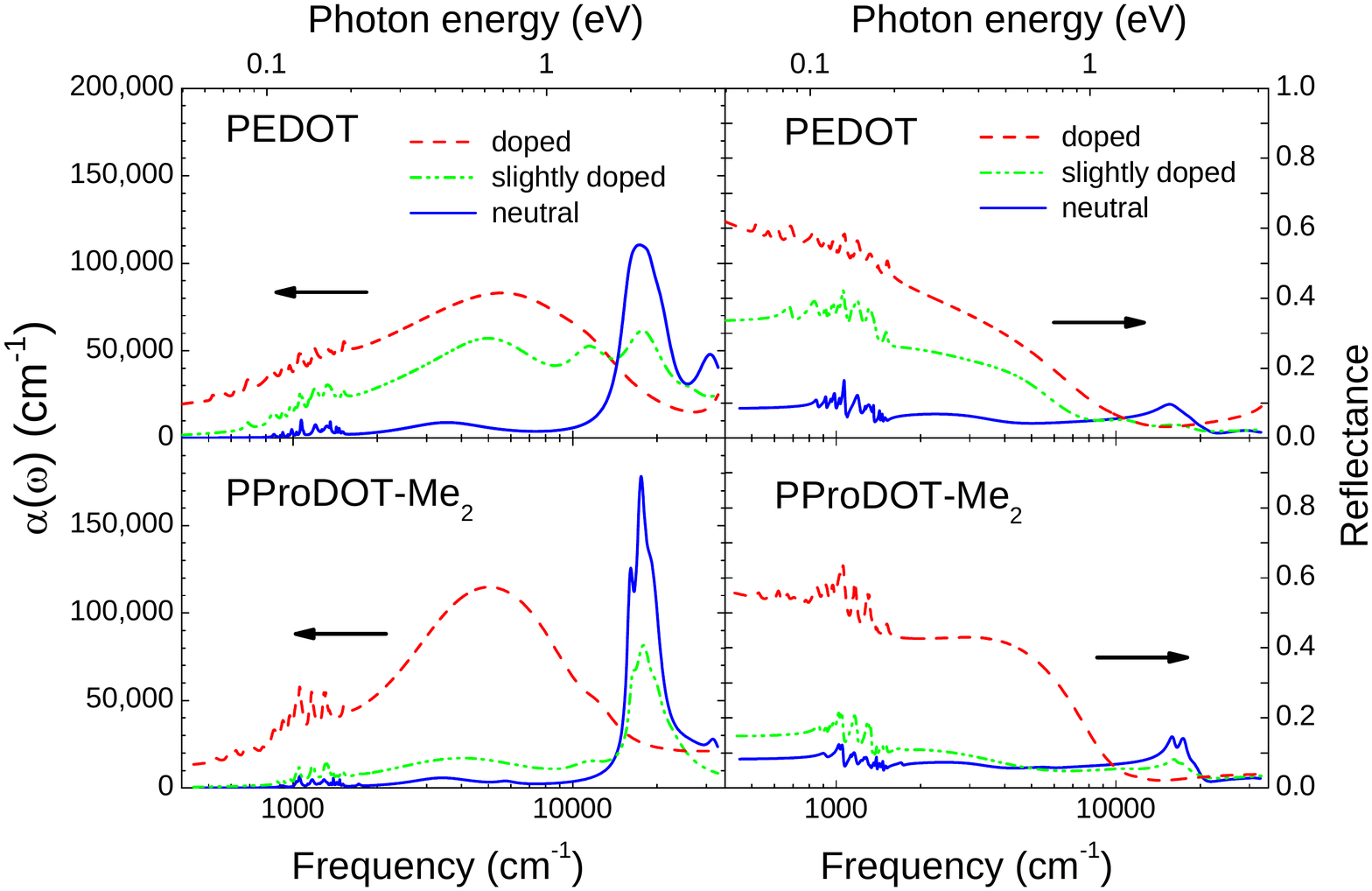}}%
 \vspace*{-0.5 cm}%
\caption{(Color online) The absorption coefficient (left panels) and reflectance (right panels) of PEDOT and PProDOT-Me$_{2}$.}
\label{Absorp}
\end{figure}

Earlier we characterized 
two thiophene-derivative conjugated polymers,
poly(3,4-ethylene\-dioxy\-thiophene) (PEDOT) and
poly(3,4-dimethyl\-propylene\-dioxy\-thiophene) (PProDOT-Me$_2$),
as thin films on
indium-tin-oxide/glass substrates.\cite{edot-ito} Because these
materials in their undoped (non-conducting) state are rather air
sensitive, we were able only to study the heavily doped conductive
phase, the undoped insulating phase, and one intermediate,
lightly-doped, phase. The reflectance data were fit to a
Drude-Lorentz (free carrier and harmonic oscillator) model, from
which the optical constants could be estimated. Figure~\ref{Absorp}
shows the absorption coefficients and bulk reflectance of these
polymers obtained from the analysis. These data are shown for photon
energies between 400 and 35,000 cm$^{-1}$ (50 meV - 4.3 eV), a range
corresponding to wavelengths between 25 $\mu$m and 290 nm. These
polymers show striking changes as they are switched between neutral
and p-doped states over the entire far-infrared-visible spectral
range. In particular, the strong $\pi$-$\pi$* transition across the
bandgap of the neutral (insulating) phase is almost completely
bleached in the doped state; at the same time a strong midgap
absorption appears. There are also changes in the vibrational
spectrum. For details we refer the reader to
Ref.~\onlinecite{edot-ito}.

In this paper we report {\it in situ} studies of electrochromism in
PEDOT and
PProDOT-Me$_2$. We
use electrochromic cells to investigate the doping-induced
sub-bandgap structure of these polymers at many doping values. The 
cells  
allow us to obtain a picture of the evolution of the sub-bandgap 
structure in these polymers as well as to observe the bleaching 
evolution of 
the $\pi$-$\pi$* transition.
We also study the effect of polymer thickness on the spectra 
and investigate strategies to minimize the 
contributions of water and gel electrolyte features.

\section{Sample Description}

\begin{figure}
\centering
 \vspace*{-0.0cm}%
 \centerline{\includegraphics[width=3.5in]{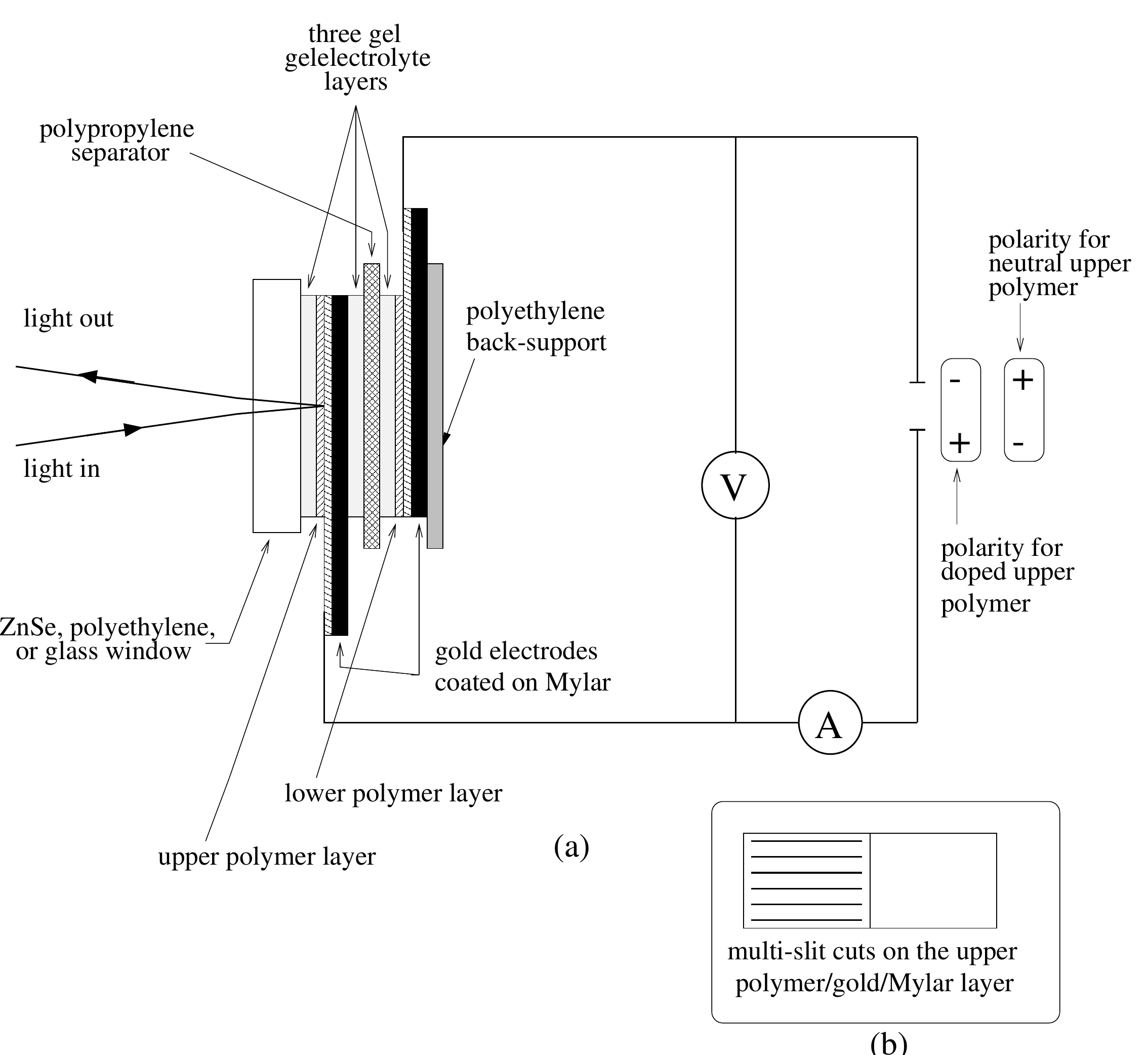}}%
 \vspace*{-0.0cm}%
\caption{(a) Schematic diagram of the electrochromic cell and circuit. (b) Top view of the upper polymer/gold/Mylar layer.}
\label{ecc2}
\end{figure}

A schematic diagram of the electrochemical or electrochromic cell and its electrical circuit is shown in Figure~\ref{ecc2}. The basic structure of the cell is an outward facing sandwich.\cite{cell} In detail, the cell consists of twelve layers, as follows (left to right in the figure): an optical window, gel electrolyte, an upper, initially p-doped, polymer film on gold/Mylar (polymer/gold/Mylar), gel electrolyte, a polypropylene separator, gel electrolyte, a lower, initially neutral, polymer film on gold/Mylar (polymer/gold/Mylar), and a polyethylene support. Part (b) of the figure shows a face-on view of the cell, illustrating the several parallel slits in the upper polymer/gold/Mylar layer. These slits allow the exchange ions in the gel electrolyte to migrate through the layer, completing the circuit.  The porous polypropylene separator serves to keep the two electrodes apart. The gel electrolyte is a complex medium consisting of four different chemical components: Acetonitrile (ACN): propylene carbonate (PC):Poly\-methyl\-meth\-acylate (PMMA):Li[N(SO$_{2}$CF$_{3}$)$_{2}$] = 70 : 20 : 7 : 3 by weight.\cite{pedot4}

All layers have important roles because all the cell components (except for the optical window and the poly\-ethylene support) make one closed electrical circuit during the {\it in-situ\/} device reflectivity measurements. However, only the upper four layers (the optical window, the gel electrolyte layer, the upper doped polymer film, and the gold film coated on Mylar) contribute to the {\it in-situ\/} device reflectivity. So those four layers are the most important for the optical response of the cell. The light path is shown in the figure.

\section{Experiment: Device reflectivity measurements}

We used a Bruker 113v FTIR spectrometer for far-infrared and midinfrared measurements (100--5,000 cm$^{-1}$ or 100--2~$\mu$m) and either a Zeiss MPM 800 microscope spectrophotometer (4,500--45,000 cm$^{-1}$ or 2200--220 nm) or a modified Perkin-Elmer 16U (3,700--45,000 cm$^{-1}$ or 2700--220 nm) for near-infrared, visible, and ultraviolet measurements.

We measured the {\it in-situ\/} device reflectivity on the top side of the cell (see Figure~\ref{ecc2}). We took care that when a thick ZnSe window was used, the focal spot of the reflectance attachment was placed on the upper active polymer surface. Even so, there was a considerable contribution ($\sim 17$\%) from the upper surface of the ZnSe.

For the {\it in-situ\/} device reflectivity measurements we adjusted the cell voltages in steps to change the doping levels of the upper active polymer layer, reading the current during the process. We used a model Lambda LL-901-OV regulated power supply as a voltage source, an HP 34401A multimeter as an ammeter, and a Fluke 70 series II multimeter as a voltmeter. We took each {\it in-situ\/} device reflectivity spectrum when we got reasonably stable values of the current and the voltage. The current in the circuit is a measure of doping uniformity; when the current  decays to about 1\% of its peak value, we judge that the doping level has stabilized. Our sign convention is that a positive voltage represents the doped polymer and a negative voltage the neutral (undoped) polymer.

\begin{figure}
\centering
 \vspace*{-1.5cm}%
 \centerline{\includegraphics[angle=0,width=4.0 in]{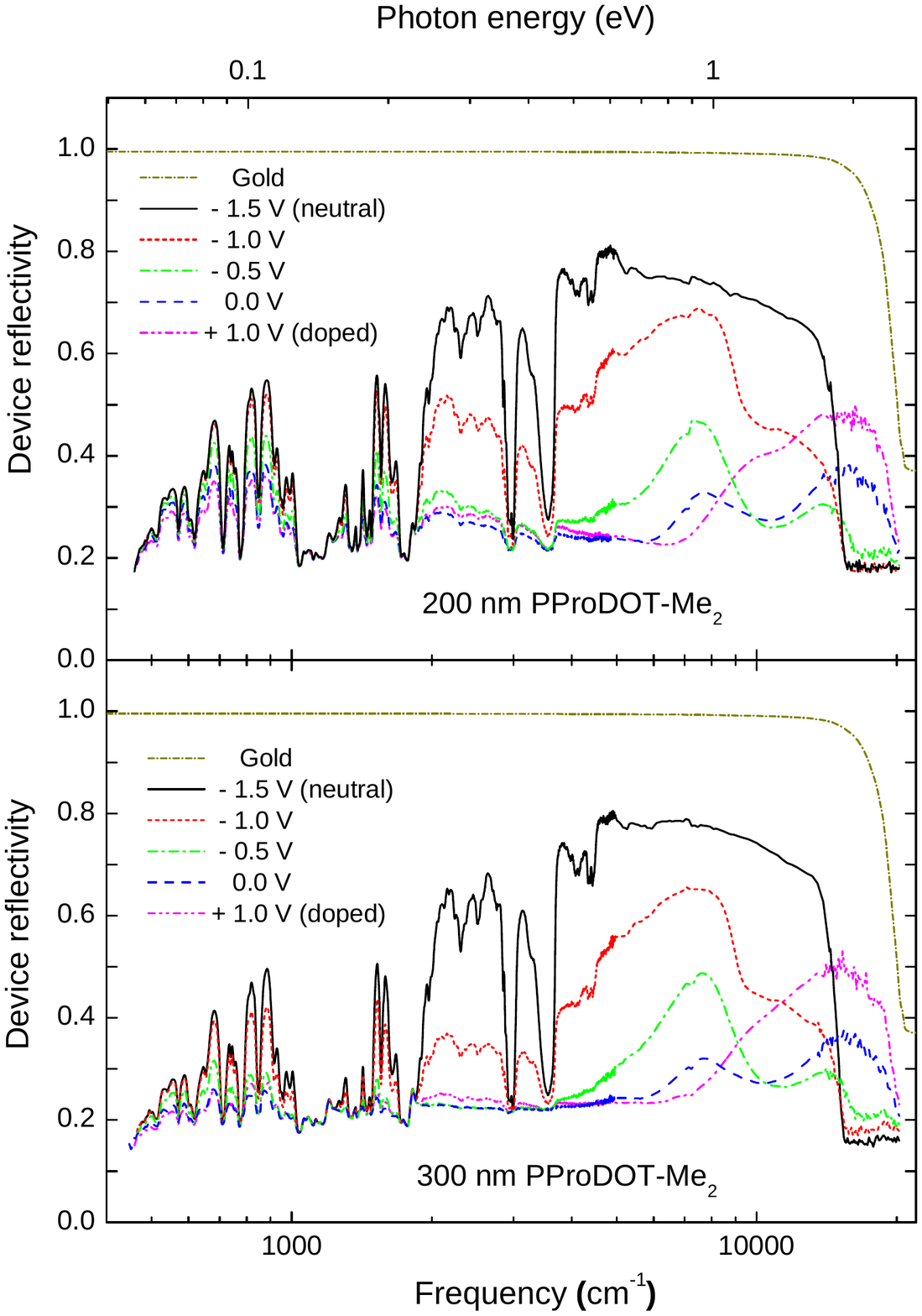}}%
 \vspace*{-2.5cm}%
\caption{(Color online) Midinfrared through visible {\it in-situ\/} device reflectivity of 200 nm and 300nm PProDOT-Me$_{2}$ films in electrochromic cells with ZnSe windows. Note the logarithmic frequency scale.}
\label{Pdmecc}
\end{figure}

\section{Results}

\subsection{Electrochromism: {\it In-situ\/} device reflectivity}

Figure~\ref{Pdmecc} shows  the reflectivity of two PProDOT-Me$_{2}$-based cells for voltages between -1.5 and +1.0 V. These cells had PProDOT-Me$_{2}$ for both the upper and lower polymer layers and employed a ZnSe window, allowing data to be taken over 450--25,000 cm$^{-1}$. The ZnSe front surface adds  about 17\% to the reflectivity in these data. The reflectivity of the gold-coated Mylar is also shown. When the polymer is neutral, (-1.5 V), the cell has the highest infrared reflectivity but the lowest visible-region reflectivity. Strong dips are seen at the O-H stretch (3500 cm$^{-1}$), the C-H stretch (2950 cm$^{-1}$), and in the fingerprint region of the infrared (600--1600 cm$^{-1}$). In contrast, when the polymer is most strongly doped, the cell has low midinfrared reflectivity and relatively high near-infrared/visible reflectivity. There is a drop around 20,000 cm$^{-1}$, just where gold's reflectance decreases.

\begin{figure}
\centering
 \vspace*{-0.5cm}%
 \centerline{\includegraphics[angle=0,width=4.0 in]{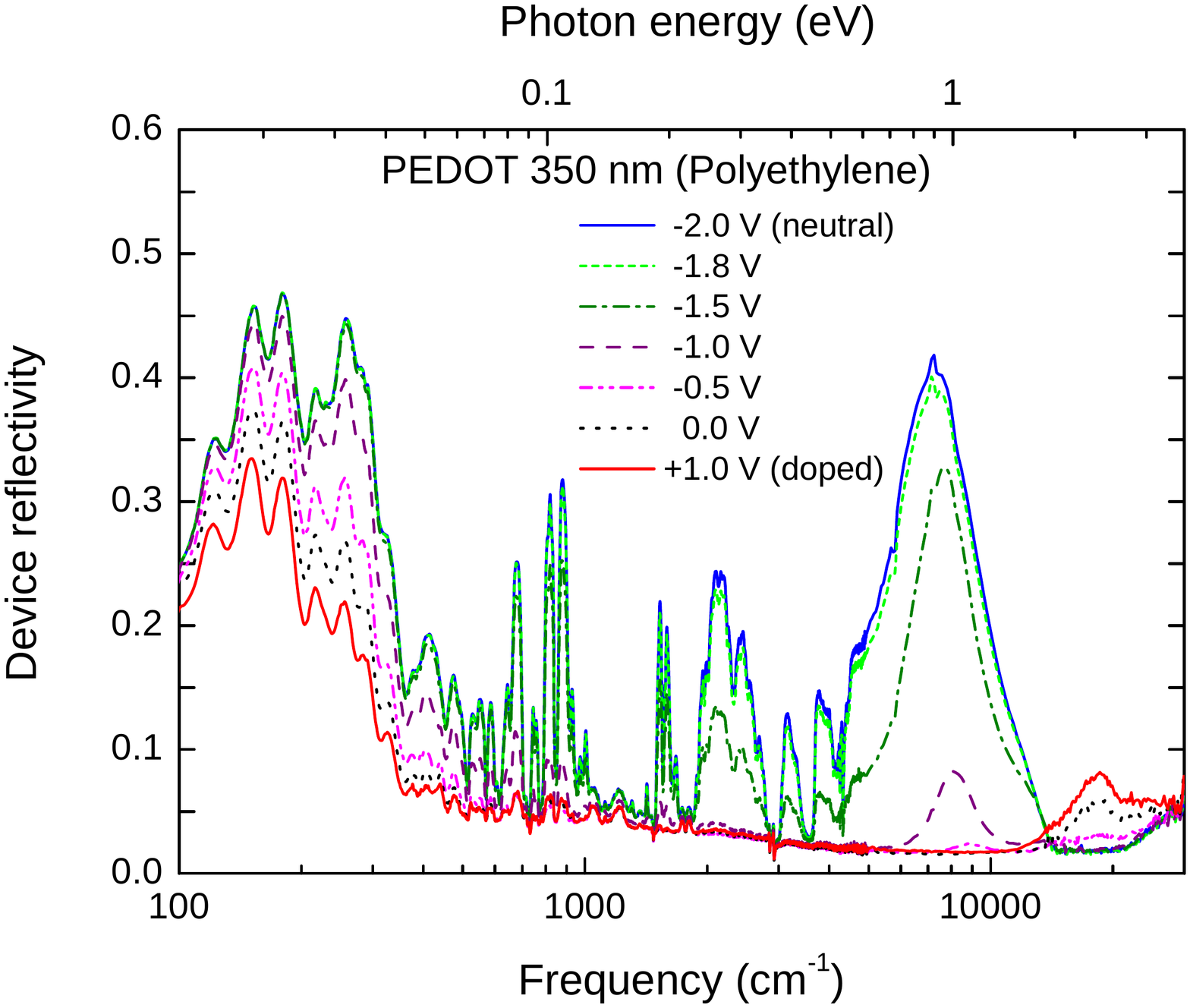}}%
 \vspace*{-0.5cm}%
\caption{(Color online) Far-infrared through visible {\it in-situ\/} device reflectivity of a 350 nm PEDOT film in an electrochromic cell with a polyethylene window. Note the logarithmic frequency scale.}
\label{Pedecc}
\end{figure}

A natural interpretation of these data is that they are dominated by the light rays which transmit through the window, gel, and polymer; are reflected by the gold; and transmit again through the polymer, gel, and window. We will shortly use this observation to analyze our results.

Figure~\ref{Pedecc} show the results for an electrochromic cell with PEDOT as the electroactive polymer. This cell used
Poly(3,6-bis(2-(3,4-ethylenedioxythiophene))-N-methylcarbazole) (PBEDOT-CZ) for the lower polymer layer. A polyethylene window was employed; polyethylene has a much lower front-surface reflection ($\sim 3$\%) and allows a much wider spectral range (100--28,000 cm$^{-1}$) than does ZnSe; however, it does contain several additional midinfrared absorption modes. The overall behavior here is similar to the cells with PProDOT-Me$_{2}$ films: neutral has higher infrared and lower visible reflectivity; heavily doped has low midinfrared but higher near-infrared/visible reflectivity.

Although different in detail, the PProDOT-Me$_2$ (Fig.~\ref{Pdmecc}) and PEDOT (Fig.~\ref{Pedecc}) electrochromic cells show qualitatively similar voltage-dependent trends. The $\pi$-$\pi*$ (or bandgap) absorption in the 15,000 to 20,000~cm$^{-1}$ (2 to 2.5 eV) is strong in the neutral phase (-1.5 V) and is bleached as the material is doped. We also observe subgap polaronic and/or bipolaronic absorption as the doping proceeds (-1.0 V, -0.5 V, 0.0 V and +1.0 V).

There is a very large difference in the {\it in-situ\/} device reflectivity between the neutral ($-1.5$ V) state and the almost fully p-doped ($+1.0$ V) state of PProDOT-Me$_2$. (See Fig.~\ref{Pdmecc})
Many strong absorption bands also appear, mostly due to the gel electrolyte. These bands limit the electrochromic response of the cell below 2000 cm$^{-1}$ (or, at wavelengths longer than 5 $\mu$m). However, at higher frequencies (between 2000 and 15,000 cm$^{-1}$ or between 5 $\mu$m and 670 nm), there is a very large difference between the neutral and doped phases. There are two strong absorption bands (the C-H stretching absorption near 3000 cm$^{-1}$ and water absorption near 3500 cm$^{-1}$) in this spectral range. In order to use the whole spectral range (2000-15,000 cm$^{-1}$), we need to remove or reduce these two peaks. An experiment to reduce these two absorption bands is discussed later in this paper.

The EDOT-based cell has a large electrochromic response over 5000 to 12,000 cm$^{-1}$, although thicker gel electrolyte and the polyethelene window limit its behavior. A significant response in the far infrared (below 300 cm$^{-1}$) is also observed.

  \subsection{Thickness-dependent electrochromism}

\begin{figure}
\centering
 \vspace*{-0.3cm}%
 \centerline{\includegraphics[width=3.5 in]{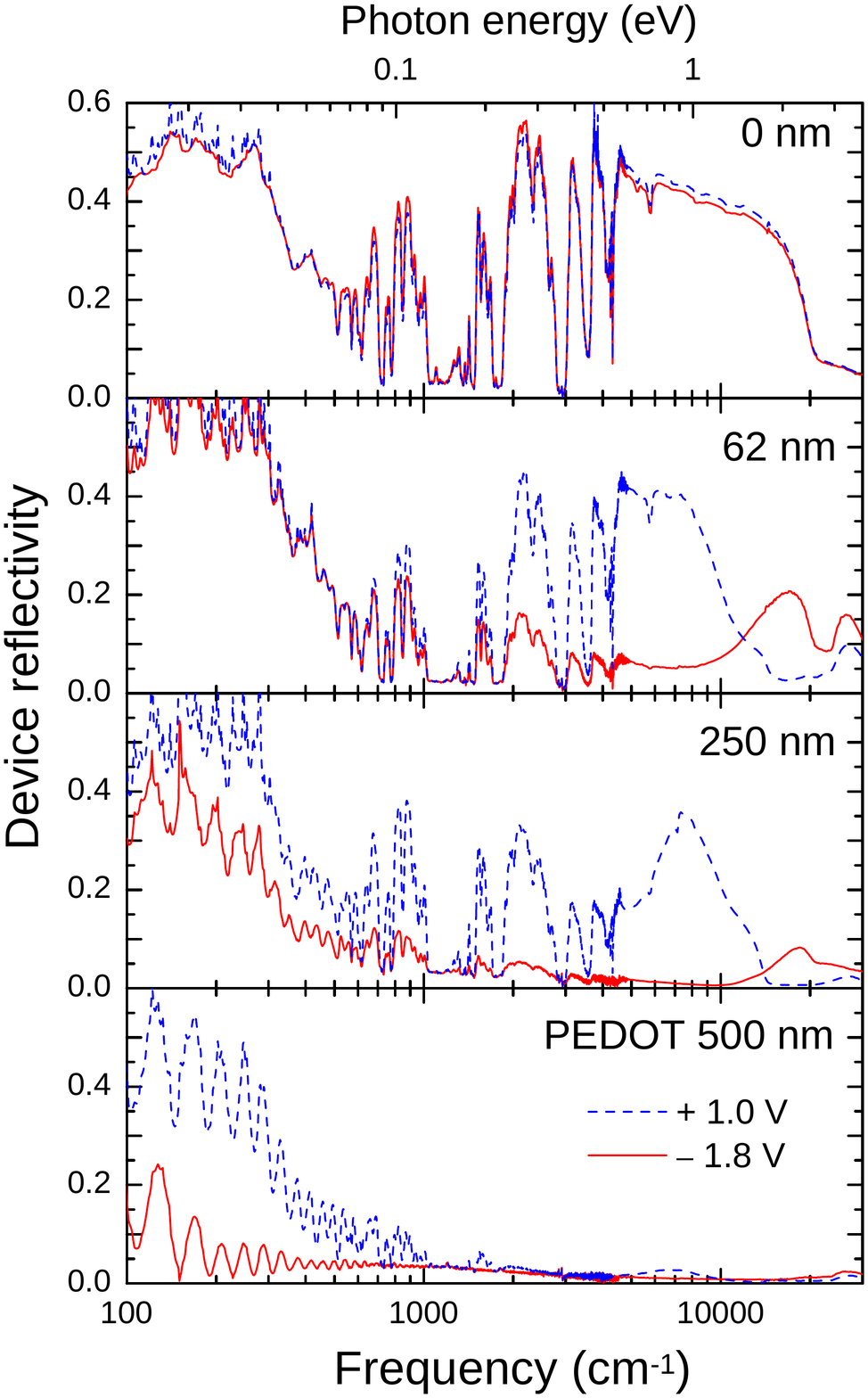}}%
 \vspace*{-0.5cm}%
\caption{(Color online) {\it In-situ\/} device reflectivity spectra of 0, 62, 250, and 500 nm thick PEDOT films in their neutral and doped states. Note the logarithmic frequency scale.}
\label{thick}
\end{figure}

We studied the dependence of the reflectivity on the thickness of the upper active polymer (PEDOT) film, preparing and measuring thicknesses of 0, 62, 250, and 500 nm for the upper active PEDOT polymer film. PBEDOT-CZ\cite{pedot4} was used as a redox polymer for the lower polymer layer in the cell. The optical windows were polyethylene.
Figure~\ref{thick} shows the {\it in-situ\/} device reflectivity data of two extreme states (neutral and doped) for 0, 62, 250, and 500 nm thick films.

 For the 62 nm film, the differences in device reflectivity of the
neutral and doped states are minimal in the low frequency range
because the neutral state of PEDOT has almost no absorption and the
doped state has low absorption (bipolaronic absorption tail) at
these frequencies. However, at higher frequencies (mid- and
near-infrared) the reflectivity in the neutral and doped states are
quite different, because the neutral state has low absorption
($\pi$--${\pi}^{*}$ transition tail) whereas the doped state has
strong absorption (bipolaronic band). In the visible, the situation
is reversed; the neutral film has a lower reflectivity than the
doped one.

These same trends are visible in the data for the 250 nm film. The
reflectivity change in the far infrared is modest, in the
midinfrared relatively large, and in the visible, reversed. The
neutral state has enough midinfrared absorption that the midinfrared
device reflectivity is smaller than in the case of the 62 nm film.

The 500 nm film shows large {\it in-situ\/} device reflectivity
changes between neutral and doped states in the far-infrared range 
because, even though the difference in absorption between the neutral and 
the doped
states is modest, the film is thick enough to show large difference
in the transmittance. (Beer's law: the transmittance of an absorbing material 
decreases 
exponentially with the thickness of the material.) However,
at high frequencies (mid- and near-infrared) the reflectivity in the
neutral and doped states is almost the same because even the low
absorption ($\pi$--${\pi}^{*}$ transition tail) of the neutral state
is strong enough to make both  neutral and doped films opaque; the
device reflectivity is governed by the prompt reflection from polyethylene
and polymer.

\section{Discussion}

\subsection{Analysis of the {\it in-situ\/} device reflectivity}

The optics of the cell is complicated. We measure the device reflectivity spectra; this does not mean that we measure the real reflectance of the polymer. The path of the reflected light is as follows. Most of the incoming light passes through the optical window (and that which reflects from the window can be subtracted from the spectra). The light also mostly passes through the thin gel electrolyte layer into the polymer film. Here there are two possible fates for the light.

In spectral regions where the polymer has low absorption (infrared in the neutral polymer and either far-infrared or visible in the doped polymer), the light passes through the polymer, bounces off the gold surface and re-passes the polymer film, the gel layer, and the optical window in reverse order. This light passes through the polymer layer twice. If there were no absorption, we should have the same reflectivity as gold (almost 100\% up to 18,500 cm$^{-1}$). However, our data always show lower reflectivity than the gold reflectivity because we always have some absorption in one or more of the three other components of our cell: electrolyte gel, polymer, and window. Thus, the data tend to be similar to transmittance spectra, plus some small reflectance.

In spectral regions where the absorption is large (visible in the case of the neutral polymer and mid infrared in the case of the doped polymer), some of the light is reflected by the polymer surface. However, the reflectance is not large (especially when the polymer is covered by gel electrolyte which acts as an index-matching region) and so most of the light enters the polymer, where it is absorbed, either on the way in to the gold layer or on the way out.

We also expect losses associated with the slits in the upper active polymer/gold/Mylar strip and losses because the surface of the polymer is not perfectly flat, especially near the slits. However, we do not see large effects from those losses in the data.

\begin{figure}
\centering
 \vspace*{-0.5cm}%
 \centerline{\includegraphics[angle=0,width=3.5 in]{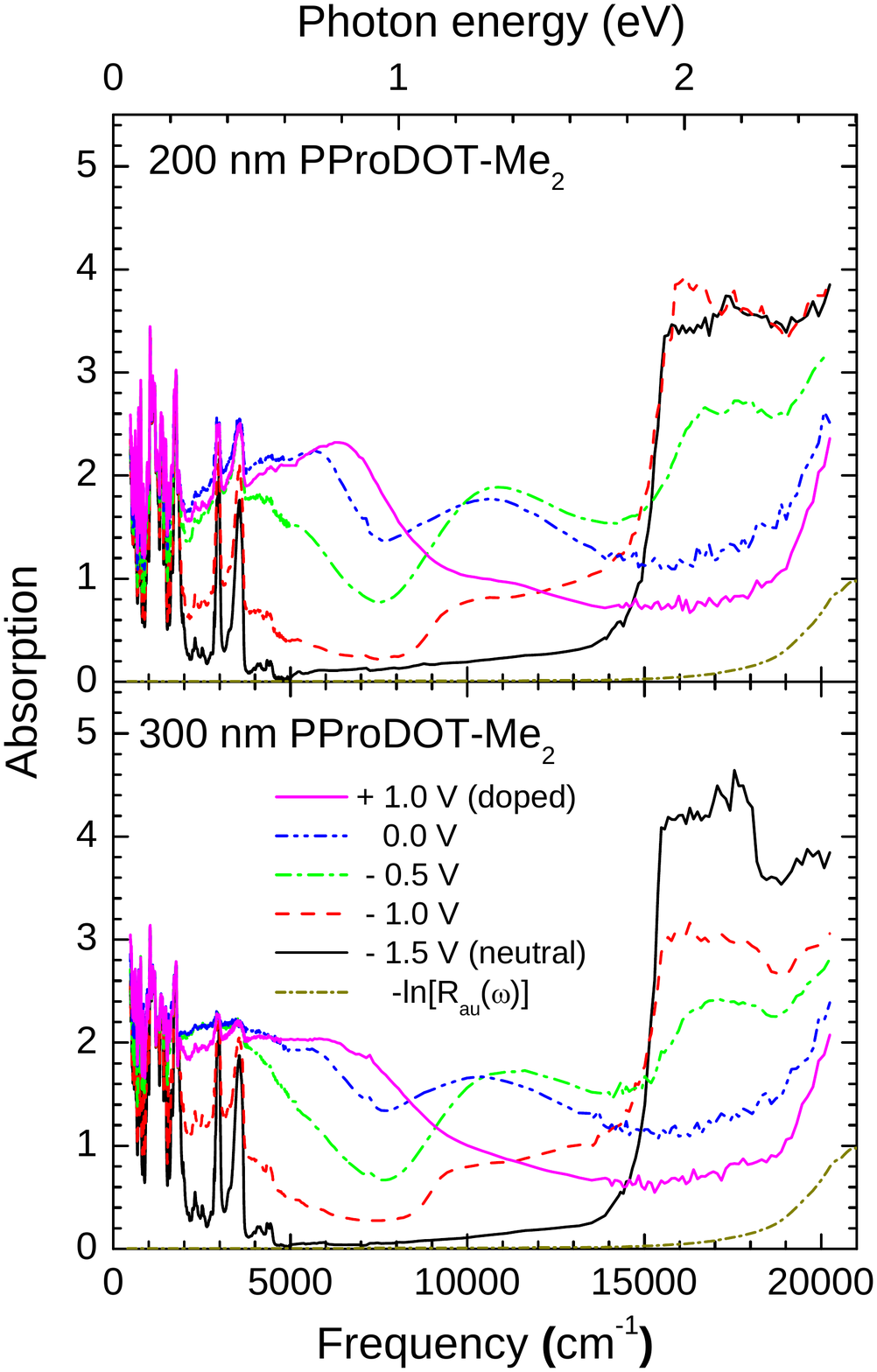}}%
 \vspace*{-0.5 cm}%
\caption{(Color online) Midinfrared through visible {\it in-situ\/} absorption of 200 nm and 300 nm PProDOT-Me$_{2}$ films in the electrochromic cell. Here the absorption is $A(\omega)-(-\ln[R_{zs}(\omega)])$.}
\label{EccAbspb}
\end{figure}

\begin{figure}
\centering
 \vspace*{-0.5cm}%
 \centerline{\includegraphics[angle=0,width=3.5 in]{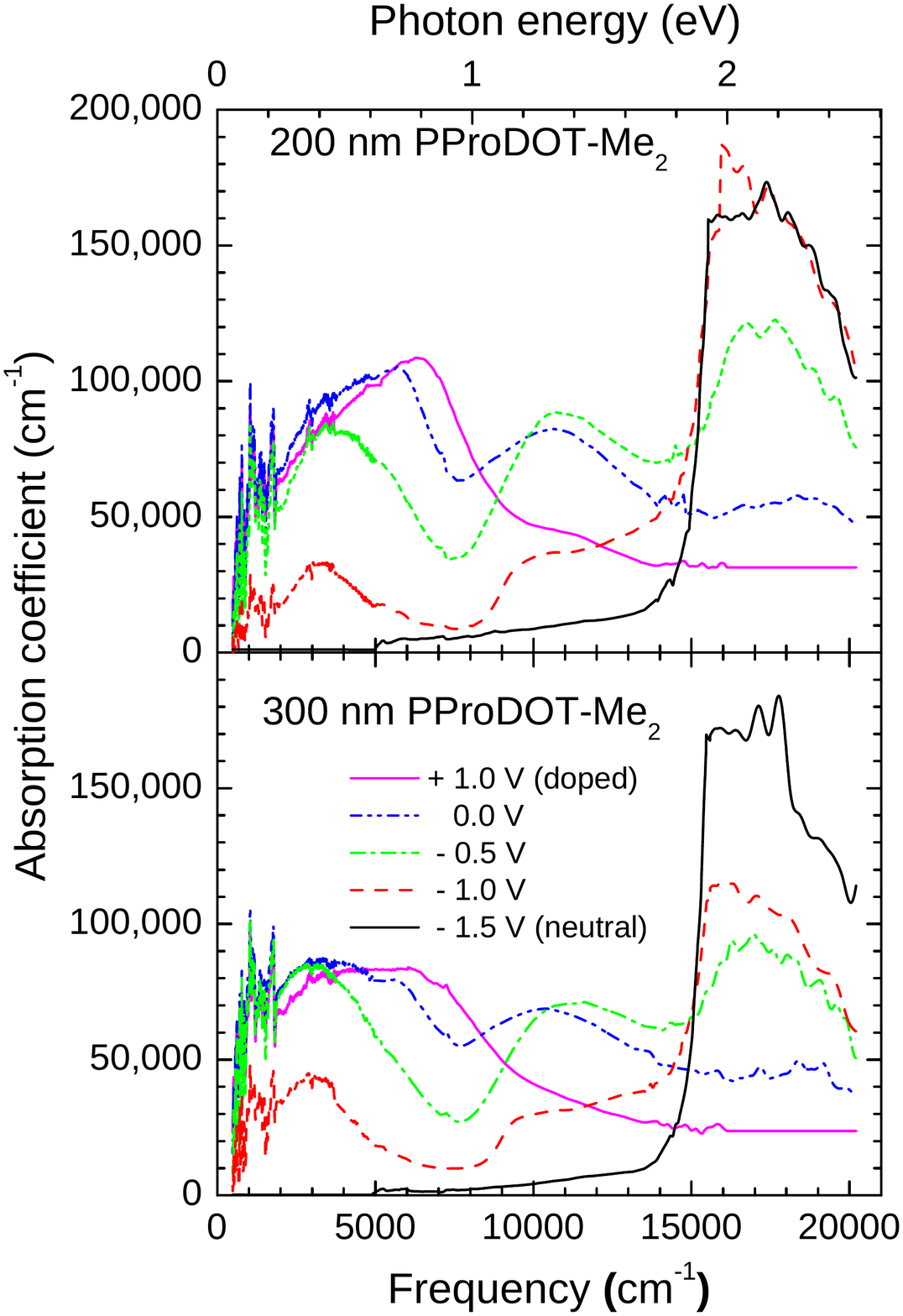}}%
 \vspace*{-0.5 cm}%
\caption{(Color online) Midinfrared through visible {\it in-situ\/} absorption coefficients of 200 nm and 300 nm PProDOT-Me$_{2}$ films with electrolyte gel and ZnSe window contributions removed. The data were corrected for thickness errors by scaling by a factor of 1.9 (200 nm film) and 2.5 (300nm film). (See text for details.) }
\label{EccAbsorp1b}
\end{figure}

\begin{figure*}
\centering
 \vspace*{-0.8 cm}%
 \centerline{\includegraphics[angle=0,width=6.0 in]{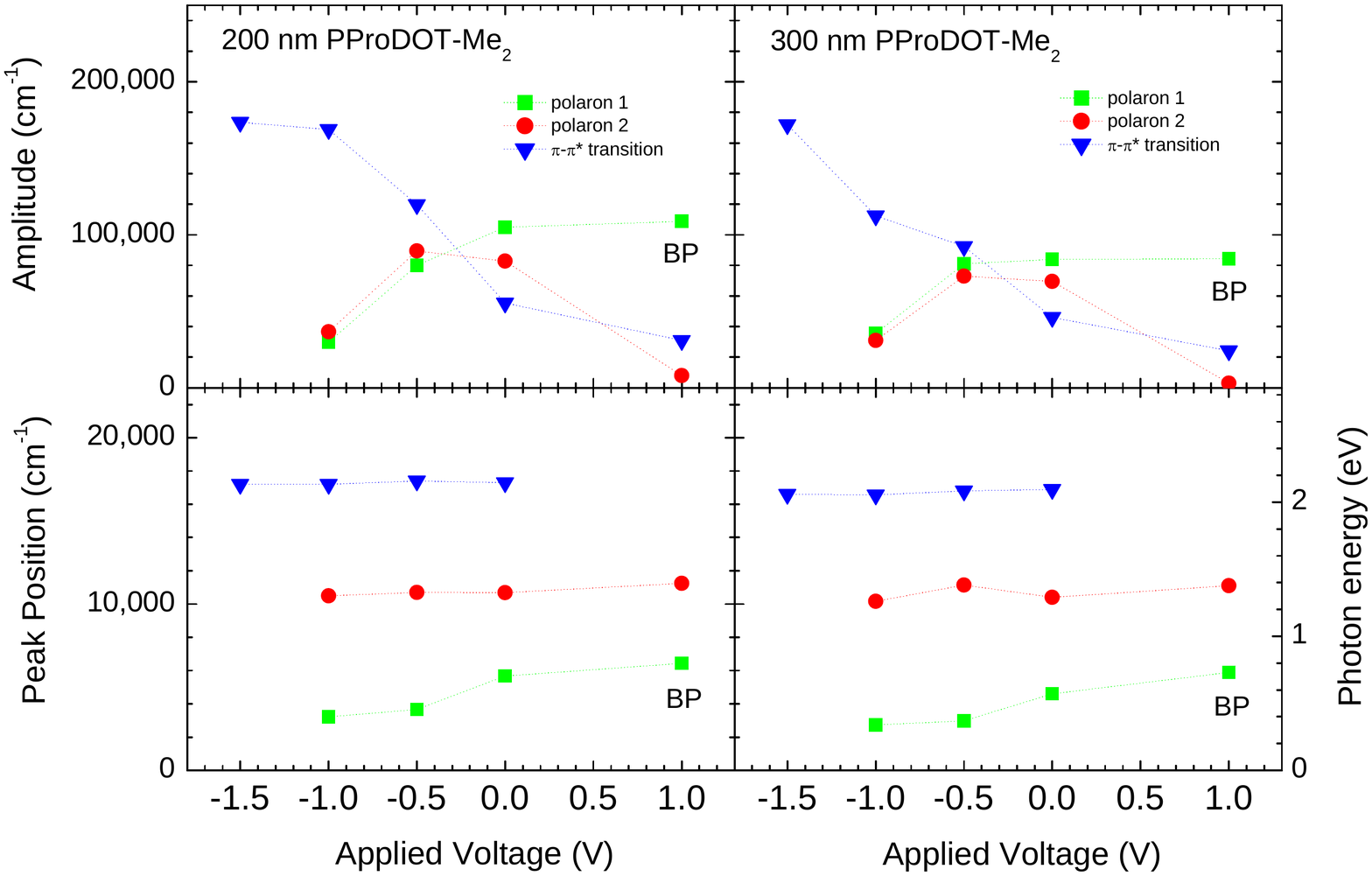}}%
 \vspace*{-1.0 cm}%
\caption{(Color online) Applied voltage/doping-dependent sub-bandgap features. The upper panels show the amplitude and the lower ones show the peak positions of the absorption bands for the 200 nm (right) and 300 nm (left) PProDOT-Me$_2$ films. BP stands for the bipolaronic mode.}
\label{subgap}
\end{figure*}

As already mentioned, the {\it in-situ} device reflectivity is dominated by light that transmits twice through the polymer layer and the gel layer, reflecting off of the gold underneath. We used this idea to extract the {\it in-situ} absorption $A(\omega)$ of those layers from the measured device reflectivity. The absorption can be written as follows:
\begin{equation}
A(\omega) = -\ln[R(\omega) - R_{zs}(\omega)]
\end{equation}
where $R \equiv R_{zs} + T_{zs} T_{gel} T_{poly} R_{au} T_{poly} T_{gel} T_{zs}$ is the device reflectivity. Here, $T_{gel}$, $T_{poly}$ and $T_{zs}$ are the transmittance of the gel layer, polymer layer and ZnSe window, respectively. $R_{au}$ and $R_{zs}$ are the reflectance spectra of gold and ZnSe. Multiple internal reflections in the window and other layers are neglected. Before calculating the absorption we subtract the parasitic reflectivity of the window, $R_{zs}(\omega)$, from the measured spectra. We display the resulting absorption, $A(\omega)$, and
$-\ln[R_{au}(\omega)]$ in Fig.~\ref{EccAbspb}. Note that the spectra contain contributions from the polymer, window, electrolyte gel, and gold electrode. The spectra show electronic transitions characteristic of the neutral ($\pi$--${\pi}^{*}$), intermediate doped (polaronic), and heavily doped (bipolaronic) polymer.\cite{edot-ito} For $-1.5$ V cell voltage, there is a very sharp $\pi$--${\pi}^{*}$ transition edge at 15,000 cm$^{-1}$ (1.9~eV). For $- 1.0$ V, intermediate features appear, attributed to excitations to and between polaron levels in the polymer bandgap. These occur at 2500 cm$^{-1}$ (0.31~eV) and 9000 cm$^{-1}$ (1.1~eV). Still present is the $\pi$--${\pi}^{*}$ transition edge at 15,000 cm$^{-1}$ (1.9~eV). For $+ 1.0$ V, there is one strong absorption, a strong bipolaronic midgap absorption at 7000 cm$^{-1}$ (0.87 eV), indicating that the cell is completely doped at this voltage.

We are able to manipulate the absorption spectra of the 200 nm and 300 nm PProDOT-Me$_2$ films to show the band gap and sub-bandgap features, removing the electrolyte gel, window absorption, and gold layer contributions to the absorption spectra. The manipulation was done in the following way. First, we divided out the gold reflectance, $R_{au}$. Then, because we already know that the undoped neutral polymer has minimal absorption below 5000 cm$^{-1}$,  the absorption features below 5000 cm$^{-1}$ are attributed to the electrolyte gel and ZnSe window. We also knew that there is a range with a constant absorption for the doped polymer above 16,000 cm$^{-1}$ (see Fig.~\ref{Absorp}). The additional absorption above 16,000 cm$^{-1}$ comes from the ZnSe window. So we can subtract the electrolyte gel and ZnSe window contributions from the absorption spectra to get the absorption of the polymer alone at various doping stages. We describe this manipulation with equations as follows:
\begin{eqnarray}
A'(\omega) &\equiv& A(\omega)-(-\ln[R_{zs}(\omega)])-(-\ln[R_{au}(\omega)]) \nonumber \\
&=& A_{gel}(\omega)+A_{poly}(\omega)+A_{zs}(\omega) \nonumber \\
&\equiv& \alpha_{gel} \:2\:d_{gel} + \alpha_{poly} \:2\:d_{poly}+\alpha_{zs}\:2\:d_{zs} \nonumber \\
\alpha_{poly}(\omega) &=& \frac{A'(\omega)-\alpha_{gel}(\omega)\: 2\:d_{gel}-\alpha_{zs}(\omega)\:2\:d_{zs}}{2\:d_{poly}}
\end{eqnarray}
where $A'(\omega)$ is the total absorption of the gel layer, polymer layer and ZnSe window. $A_{gel}$, $A_{poly}$ and $A_{zs}$ are the absorption of the gel layer, polymer layer, and ZnSe window respectively. $\alpha_{gel}(\omega)$, $\alpha_{poly}(\omega)$ and $\alpha_{zs}(\omega)$ are the absorption coefficients of the gel, polymer and ZnSe respectively. $d_{gel}$, $d_{poly}$ and $d_{zs}$ are the thicknesses of the gel layer, polymer layer and ZnSe window respectively. The resulting absorption coefficient spectra, $\alpha_{poly}(\omega)$, are displayed in Fig.~\ref{EccAbsorp1b}. Here we had to multiply by a scale factor so that the magnitude of the absorption coefficient agreed with the data in Fig.~\ref{Absorp}. The scaling factors are 1.9 and 2.5 for the 200 nm and 300 nm films, respectively. The cause of the disagreement in magnitude is probably that the actual film thicknesses were less than what was estimated from the electropolymerization time and current. The discrepancy between real and estimated thicknesses is larger for the thicker PProDOT-Me$_2$ film.

There is remaining sharp structure below 2000 cm$^{-1}$, due to imperfect removal of the electrolyte gel contribution. However the polaronic and bipolaronic sub-bandgap absorption peaks\cite{edot-ito} are clearly seen, because these features are much broader than the gel features. In principle, after the thickness adjustment mentioned above, the absorption coefficients of the 200 nm and 300 nm PProDOT-Me$_2$ films should be the same at a given wavelength and doping. Those two sets of absorption coefficient spectra of the 200 nm and 300 nm PProDOT-Me$_2$ are indeed similar, showing a similar doping/voltage dependent trend. There are three peaks (two for the polaronic absorptions and one for the $\pi$--$\pi*$ transition) at low doping levels ($V$ = $-1.0$ V, $-0.5$ V and $0.0$ V) and one single peak (the bipolaron absorption) at the highest doping level ($V$ = $+1.0$ V).

We determined the peak positions and amplitudes of the broad polaronic and bipolaronic absorption bands and plot them as functions of the applied voltage in Fig.~\ref{subgap}. As doping increases (voltage increases), the strength of the $\pi$--$\pi*$ transition decreases while the strengths of both polaronic bands increase. Interestingly, while the lower frequency polaronic mode (P1)  becomes the bipolaron band (BP) as doping continues to increases\,  the higher frequency polaronic mode (P2) disappears. The peak position of P1 increases with doping, whereas that of P2 does not change much.

 \subsection{Absorption coefficient from thickness dependent measurements}

\begin{figure}
\centering
 \vspace*{-0.5 cm}%
 \centerline{\includegraphics[width=3.3 in]{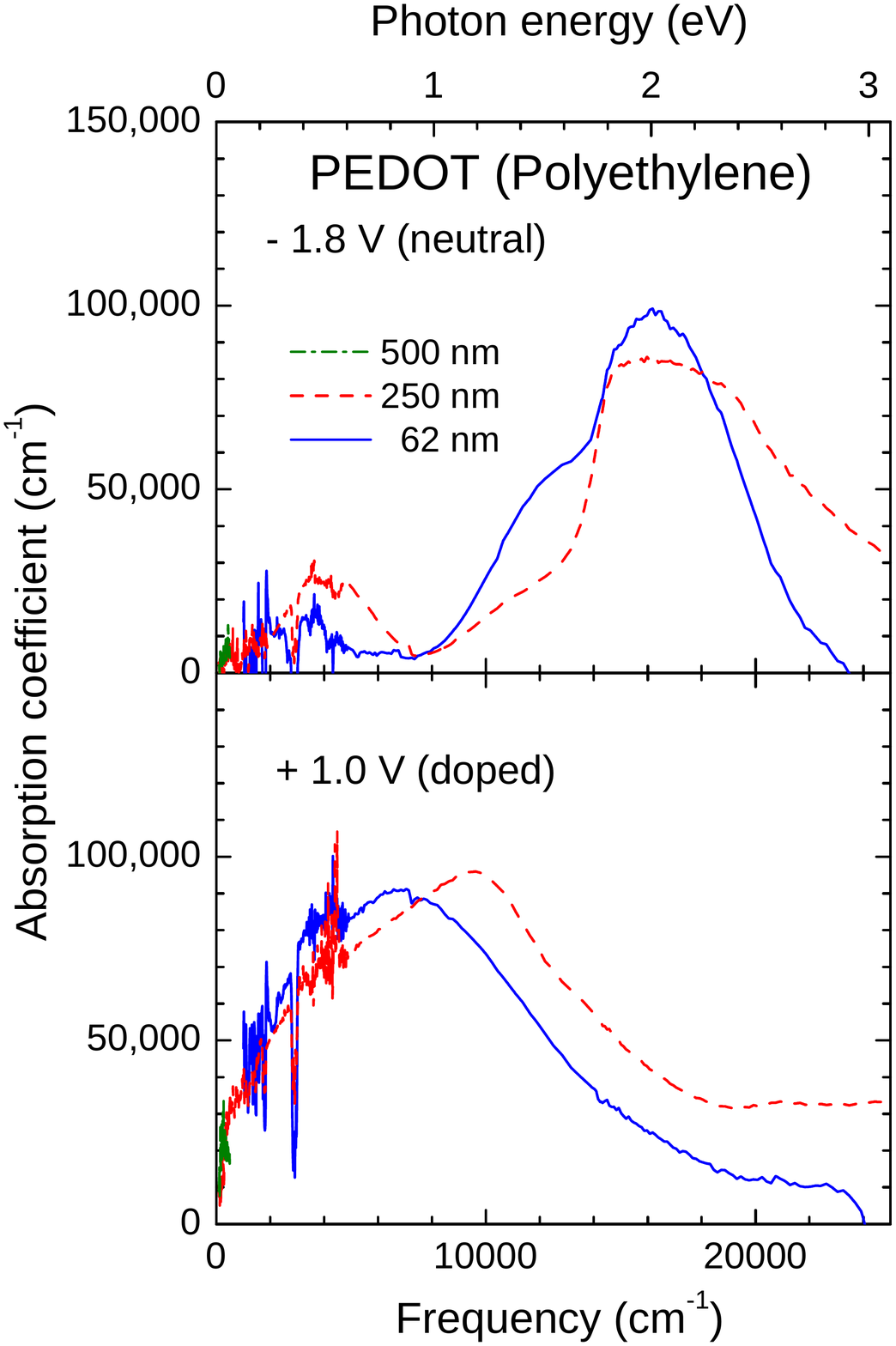}}%
 \vspace*{-0.5 cm}%
\caption{(Color online) Absorption coefficients of 0, 62, 250, and 500 nm thick PEDOT films at two doping stages: neutral (upper panel) and doped (lower panel). We also adjusted the amplitude by multiplying a scaling factor; 0.5, 1.1 and 1.0 for 62nm, 250 nm and 500nm films, respectively. (See text for details.)}
\label{thickAnalyb}
\end{figure}

We applied the same procedure to extract the absorption coefficients of the PEDOT films whose device reflectivity is shown in
Fig.~\ref{thick}. The polyethylene window and electrolyte gel were removed reasonably well. If we look at device reflectivity in Fig.~\ref{thick}, it is clear that 62 nm device reflectivity gives no information below 500 cm$^{-1}$ and the 500 nm device reflectivity gives very little information above 1000 cm$^{-1}$. The 250 nm device reflectivity seems useful over the whole range. To get the absorption coefficient of PEDOT, first we multiply the absorption coefficient of 250 nm film by a scale factor (1.1) to make it agree with the absorption coefficient of PEDOT in Fig.~\ref{Absorp}. Then, we take the adjusted absorption coefficient of the 250 nm film as the correct one. In the high frequency range, we scaled the 62 nm film to agree with the adjusted absorption coefficient of 250 nm film. We used the same scale factor (0.5) for both neutral and doped 62 nm film. We show 62 nm data only above 1000 cm$^{-1}$. In the low-frequency range, we scaled the 500 nm film in the same way, finding that the scale factor for the 500 nm film is 1.0. We report 500 nm data only below 1500 cm$^{-1}$. The resulting absorption coefficients are shown in Fig.~\ref{thickAnalyb}. The upper panel shows the absorption coefficient of the neutral films and the lower one shows that of doped films. It seems that the neutral state is not completely neutralized since there are subgap features in the absorption spectra. Nevertheless, the $\pi$-$\pi^*$ transition peak is evident around 16,400 cm$^{-1}$ in the upper panel and the bipolaronic absorption peak in the lower panel.

\section{Device Characterization}

   \subsection{Switching time}

\begin{figure}
\centering
 \vspace*{-1.5cm}%
 \centerline{\includegraphics[width=3.7 in]{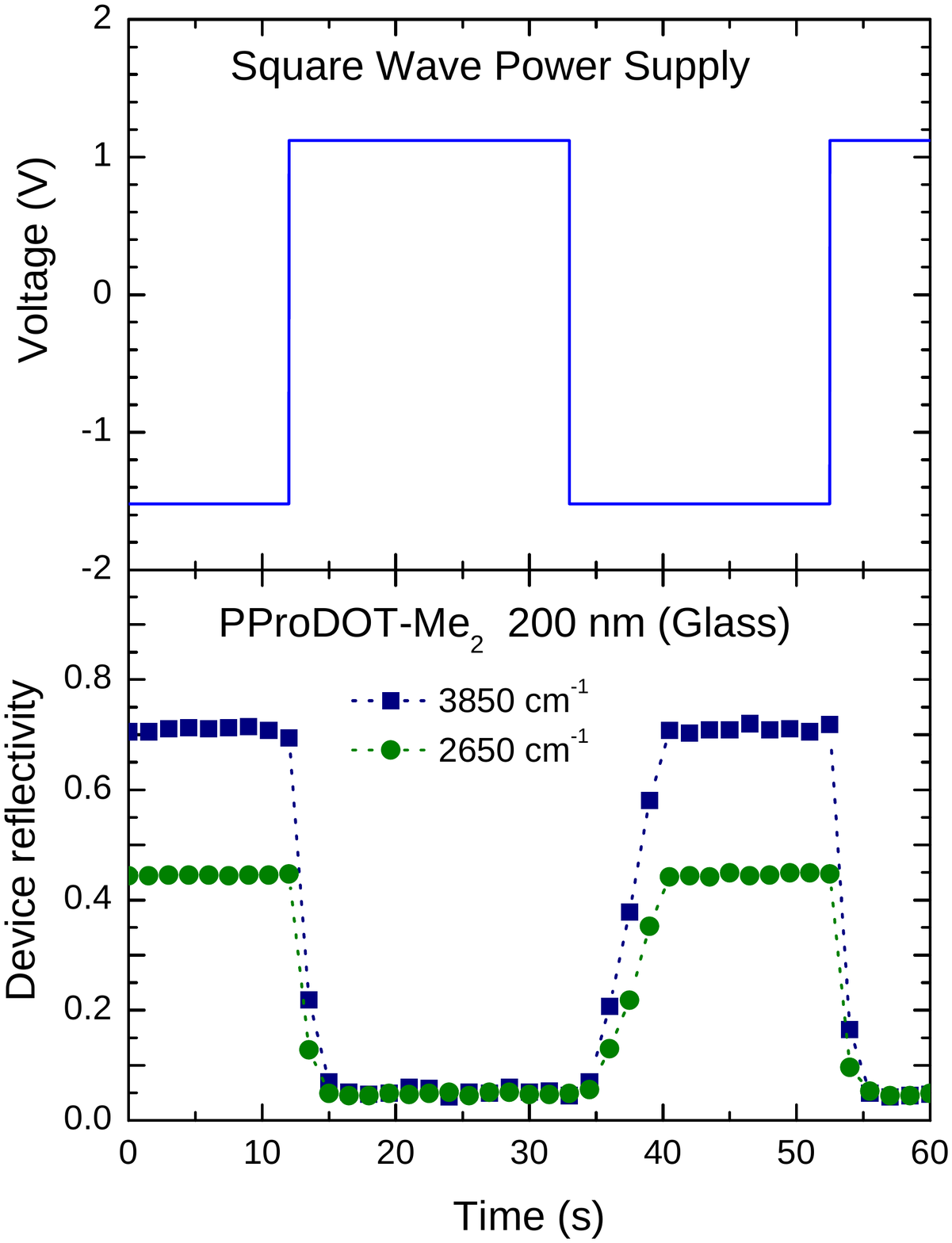}}%
 \vspace*{-1.5cm}%
\caption{(Color online) Square wave potential and infrared response on switching between neutral and doped states.}
\label{sw1}
\end{figure}

The switching (response) time of the cell can be defined as the time needed to change from the neutral state to the completely doped state or vice versa. To study the switching time we used a power supply which can give a square-wave potential. We could control the period and the amplitude of the potential. We used a PProDOT-Me$_{2}$ electrochromic cell with 200~nm PProDOT-Me$_{2}$ films for both the upper and lower polymer layers. The measurement was performed with the Bruker 113v spectrometer, using 40 scans over 1.5 sec to produce a set of spectra showing the time-evolution of the reflectivity. Figure~\ref{sw1} shows the reflectivity at two frequencies (2,650 and 3,850 cm$^{-1}$) as a function of the time. We selected 2,640 and 3,850 cm$^{-1}$ because the absorption of the gel electrolyte is small at these frequencies. The color change during the doping and dedoping processes started near the slits and spread over whole surface of the upper polymer film. Thus, the switching time measured here is determined in part by the slit separation.

Figure~\ref{sw1} shows a typical cycle: the square wave potential in the upper panel and curves of device reflectivity vs.~time in the lower panel for the two fixed infrared frequencies. After the supply has toggled from one voltage to the other we see the response of the cell. Switching from the neutral to the completely doped state (p-doping) requires about 3 seconds; switching from the completely doped to the neutral state (p-dedoping) requires about 7.5 seconds. Similar results were reported for PEDOT.\cite{pedot3} When we have a polymer film on a metallic substrate in contact with anions and cations in  solution,  we may determine which ions are getting into or out of the polymer film (or which ions are the exchanging ions) by comparing the p-doping time with the p-dedoping time. If the anions are the exchange ions the p-doping time is shorter than the p-dedoping time, and vice versa for the other case. In our case the p-doping time (3 s) is shorter than the p-dedoping time (7.5 s), leading to the conclusion that the anion, i.e., Li$^{+}$, is the exchanging ion.

   \subsection{Long-term switching stability}

\begin{figure}
\centering
 \vspace*{-0.5cm}%
 \centerline{\includegraphics[width=4.0 in]{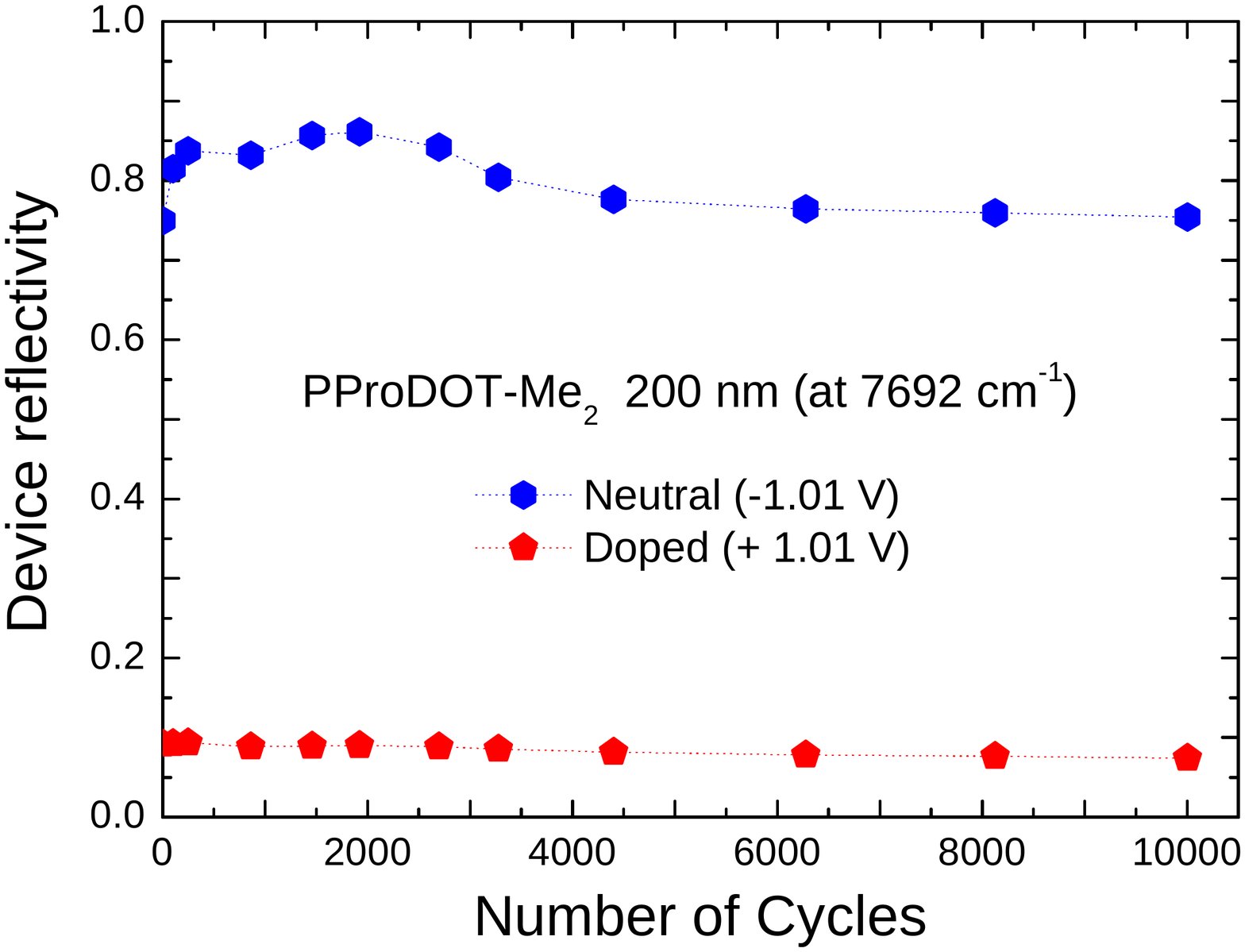}}%
 \vspace*{-0.5cm}%
\caption{(Color online) Lifetime test of the PProDOT-Me$_{2}$ electrochromic cell: device reflectivity vs.~the number of deep double potential switches.}
\label{life}
\end{figure}

To show that the observed large reflectivity changes over a broad spectral range are not due to degradation of the polymer, we did a long-term switching experiment. This experiment was performed by using the Zeiss MPM 800 microscope spectrophotometer. The experiment was done in the lab environment (300~K and in air). We used glass as the optical window in the experiment. Each measurement was performed after several deep double potential switches. During the {\it in-situ\/} device reflectivity scan, we kept the voltage constant (either $-1.01$ V or $+1.01$ V). The period of one deep double potential switch was 47 seconds: $-1.01$ V for 24 seconds, and $+1.01$ V for 23 seconds. The last data point of the measurement was taken after 10,000 double switches. It took about five and half days to conduct the experiment.

Figure~\ref{life} shows a graph of device reflectivity at 7,692 cm$^{-1}$) vs.~the number of double switches. We selected 7,692 cm$^{-1}$ because there is a large difference between neutral and doped states at this frequency; see Fig.~\ref{Pdmecc}. The result confirms that PProDOT-Me$_{2}$ in the cell is a very stable material for the long-term redox switching: after 10,000 deep double potential switches the cell shows only about a 10\% drop in the reflectivity difference between neutral and doped phases, i.e., PProDOT-Me$_{2}$ is still electrochemically active.

Factors important for the switching stability of the cell include the thickness of the polymer film, the area of the film, the solvent in the gel electrolyte,\cite{life} and the cell voltage. Among these factors, the cell voltage is the most important. We performed several lifetime test experiments using cell voltages of $- 1.51$ V and $+ 1.0$ V. After just 400--500 cycles, the electrochromic performance had decreased by around 85\%. The reason is that the lower polymer layer is overoxidized by the $- 1.51$ V cell voltage. The ``overoxidation'' voltage can be found by measuring the {\it in-situ\/} DC conductivity; the DC conductivity initially increases with the voltage or doping level but at some voltage we observe a maximum in the DC conductivity. Any voltages above this voltage are overoxidation voltages.\cite{pani} Our lifetime test experiment finds that the overoxidation voltage is above $+ 1.0$ V but below $+1.5$ V. The overoxidation is irreversible and consumes significantly more charge than the reversible oxidation.\cite{overox} It is generally believed that nucleophiles such as H$_{2}$O or OH$^{-}$ attack highly oxidized thiophene rings, leading to a breakdown of the polymer backbone $\pi$-conjugation.\cite{overox1}

\subsection{Uniformity of doping}

\begin{figure}
\centering
 \vspace*{-1.0cm}%
 \centerline{\includegraphics[width=3.8 in]{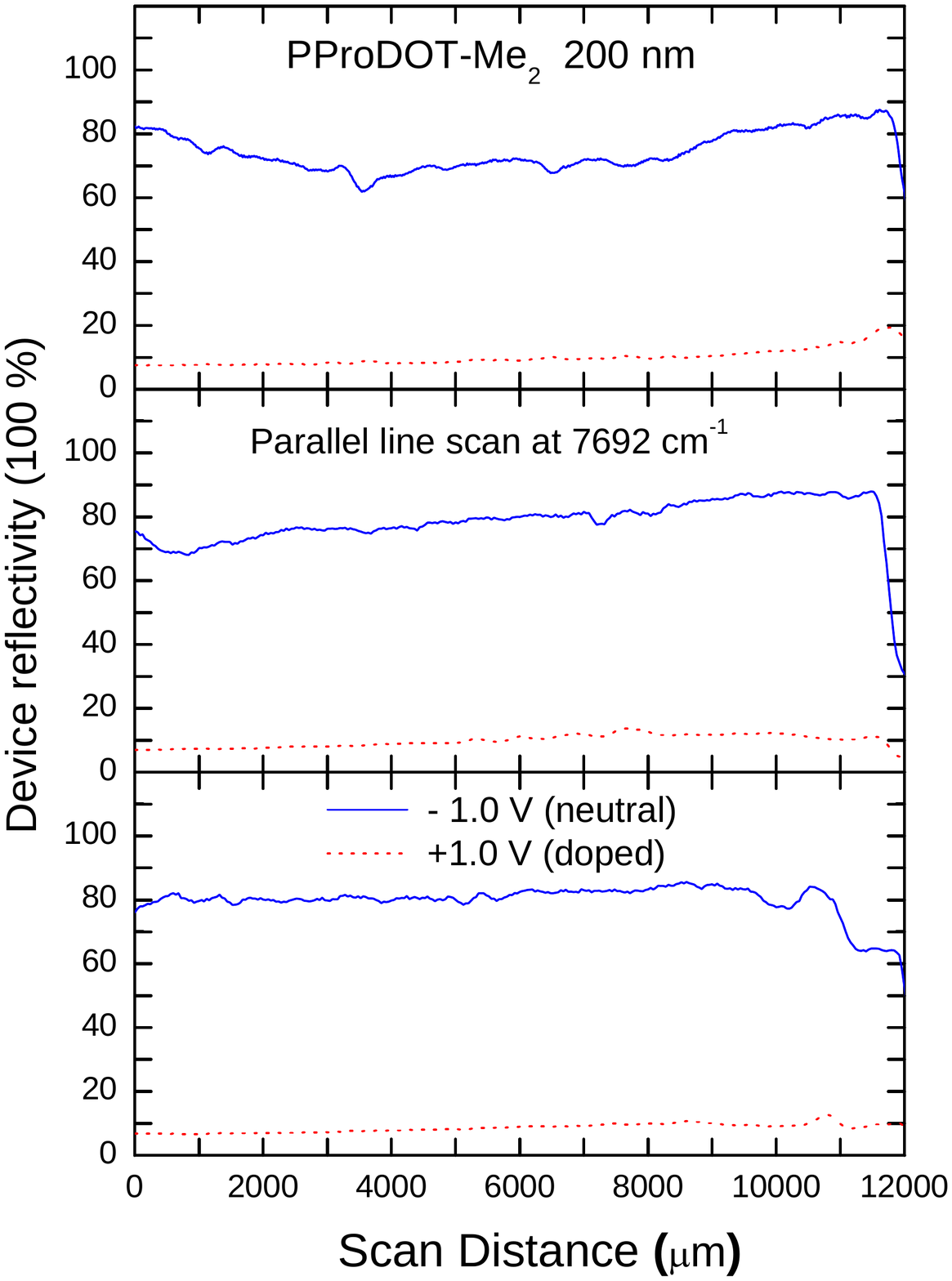}}%
 \vspace*{-1.5cm}%
\caption{(Color online) Line scans parallel to the slits after 10,000 deep double potential switches.} \label{para}
\end{figure}

\begin{figure}
\centering
 \vspace*{-1.0cm}%
 \centerline{\includegraphics[width=3.8 in]{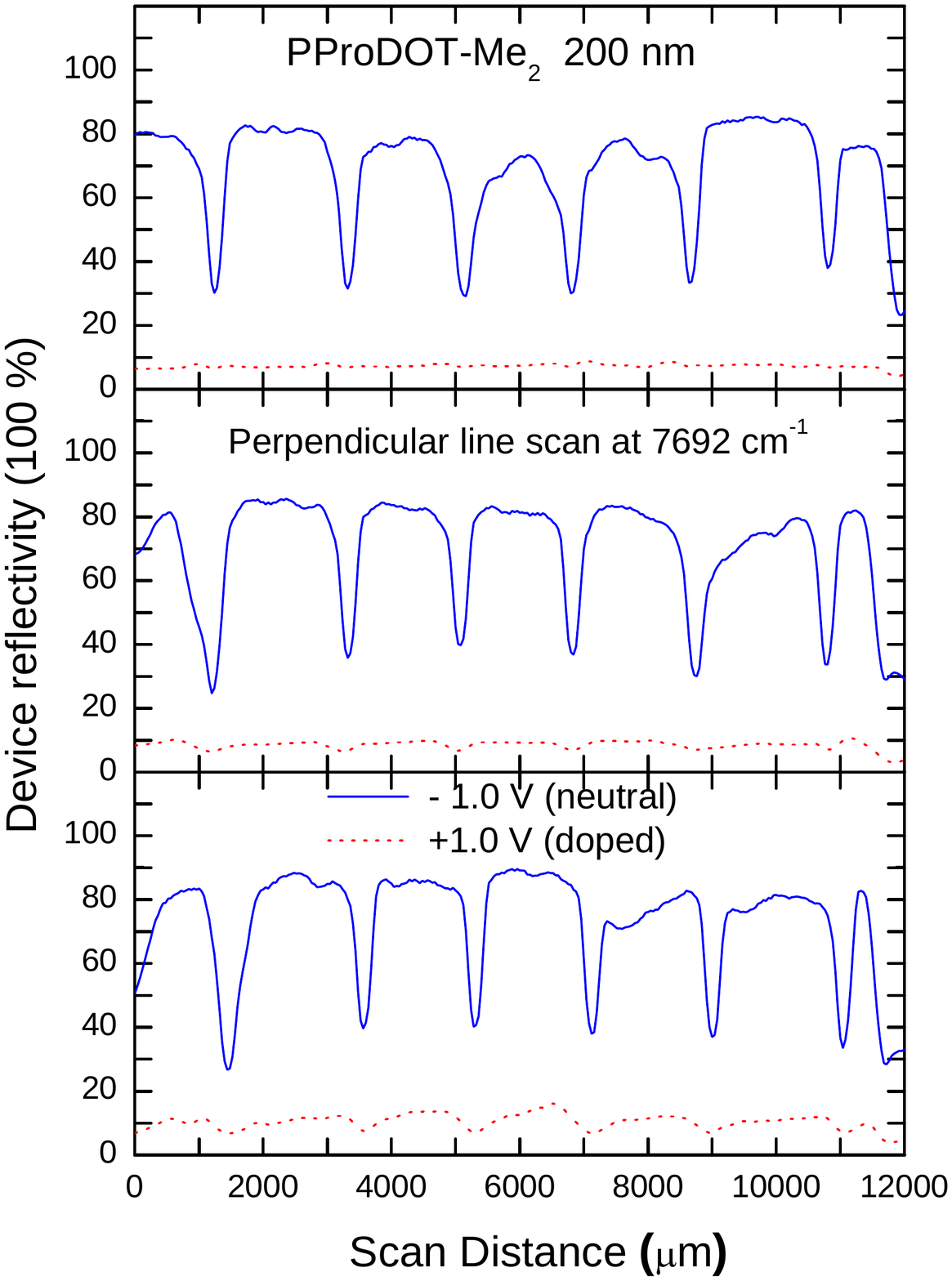}}%
 \vspace*{-1.5cm}%
\caption{(Color online) Line scans perpendicular to the slits after 10,000 deep double potential switches.} \label{perp}
\end{figure}

The Zeiss MPM 800 microscope spectrophotometer uses a small focal spot (200$\times$200 {{$\mu$}m}$^{2}$) relative to the size of the polymer surface (1.27$\times$1.27 cm$^{2}$). Another check of the stability of the cell is to look for variations in the device reflectivity from various points on the upper polymer surface. After 10,000 deep double potential switches we used the spatial line scan function of the Zeiss MPM 800 microscope spectrophotometer to measure six line scans on the upper polymer surface at a fixed infrared frequency (${\nu}=7692$~cm$^{-1}$). Three equally spaced ($\sim$ 4 mm) scan lines were made parallel to the slits on the upper polymer/gold/Mylar and three perpendicular to the slits. The step size of the scan was 40 $\mu$m. The results of the scans are shown in Figs.~\ref{para} and \ref{perp}. In Fig.~\ref{perp} the several sharp valleys correspond to the slits on the upper polymer/gold/Mylar; the distance between slits is $\sim$2 mm. Except for the dips at the slits, the variation is about $\pm10$\% of the average reflectivity.

\subsection{Removal of water and minimization of the gel contribution}

We investigated methods to improve the electrochromic properties of the cell in the midinfrared and near-infrared regions, where absorption features due to water and the gel electrolyte are found. We were able to remove water from the cell by preparing the electrolyte gel in an argon-filled glove box. The details of the processes are as follows. The electrolyte gel is made of a mixture of four chemicals, acetonitrile (ACN), propylene carbonate (PC), poly-methyl\-meth\-acylate (PMMA), and Li[N(SO$_2$CF$_3$)$_2$]. The ACN acts as a solvent and evaporates from the gel relatively quickly; its influence on the gel spectrum is minimized by letting the gel mix for several hours as the ACN evaporates. The gel was mixed for roughly 3-4 hours in the argon environment before it was stored. Use of this gel was restricted to the argon environment to ensure that no moisture was ever in contact with it. The argon environment was also maintained at a very low oxygen concentration, from 1.5 to 2 ppm at most times and never rising above 4 ppm. We used a 2 mm thick ZnSe window to protect the cell from the environment. The thickness of the PProDOT-Me$_2$ polymer film was 200 nm. Assembly of the cell was performed entirely within the argon environment to keep moisture out and to minimize oxygen exposure to the polymer. An airtight reflectance stage was designed to hold the electrochromic cell in the 
spectrometer.\cite{cornick}  The sample was clamped between two 
O-ring-sealed ZnSe windows. The cell was designed so the windows compressed the electrochromic
 sandwich structure, minimizing the thickness of the of electrolyte gel. 
 Electrical leads were brought into the sealed chamber to contact the electrodes 
and control the doping.

\begin{figure}
\centering
 \vspace*{-1.0cm}%
 \centerline{\includegraphics[width=3.3 in]{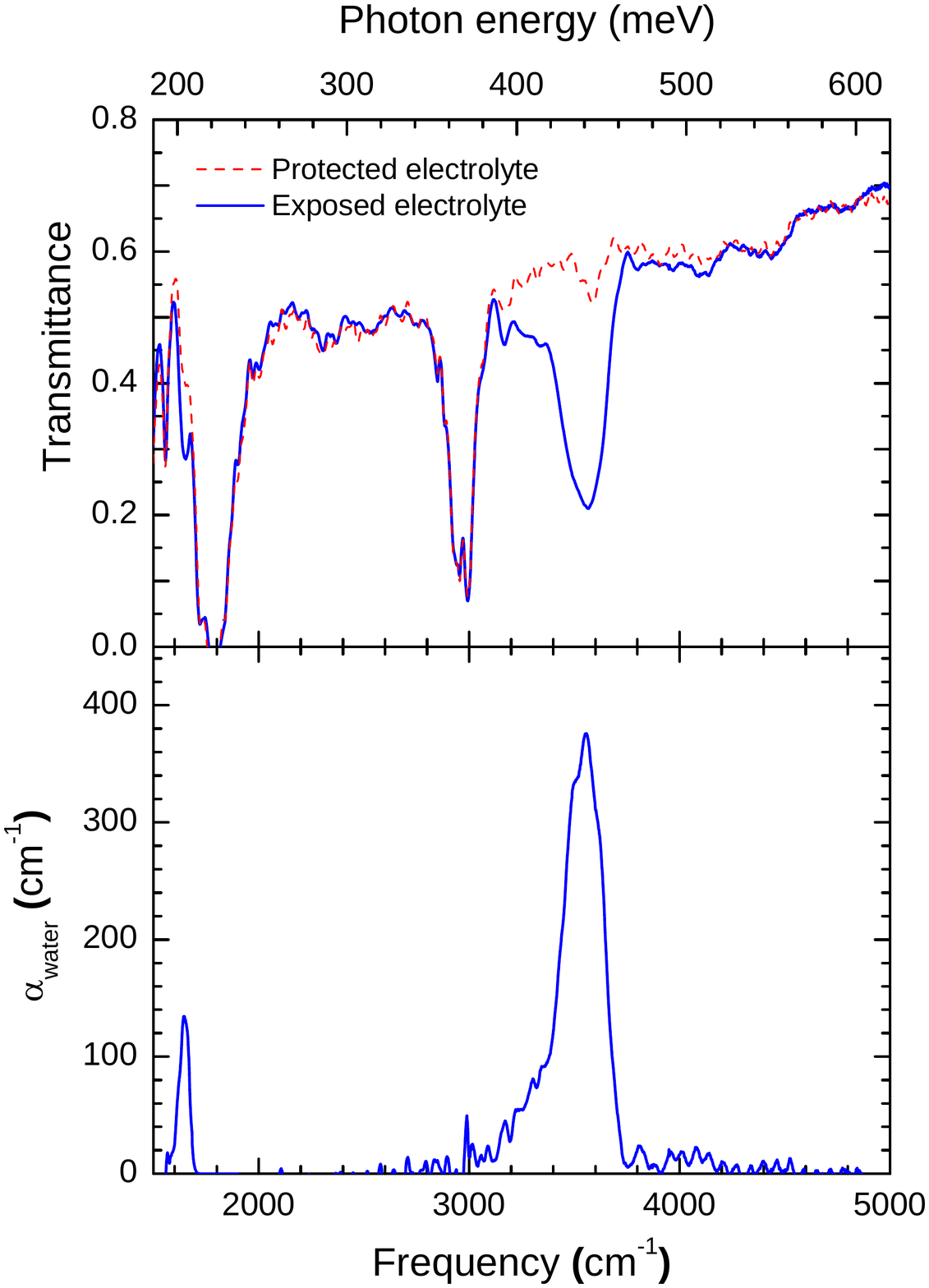}}%
 \vspace*{-1.3cm}%
\caption{(Color online) Upper panel: transmittance spectra of two 25 $\mu$m thick electrolyte gel samples, one containing water. Lower panel: absorption coefficient of the water within the gel electrolyte, calculated using the contaminated and non-contaminated gel transmittance spectra. (See text for  details.)} \label{EGel}
\end{figure}

\begin{figure}
\centering
 \vspace*{-0.7cm}%
 \centerline{\includegraphics[width=3.3 in]{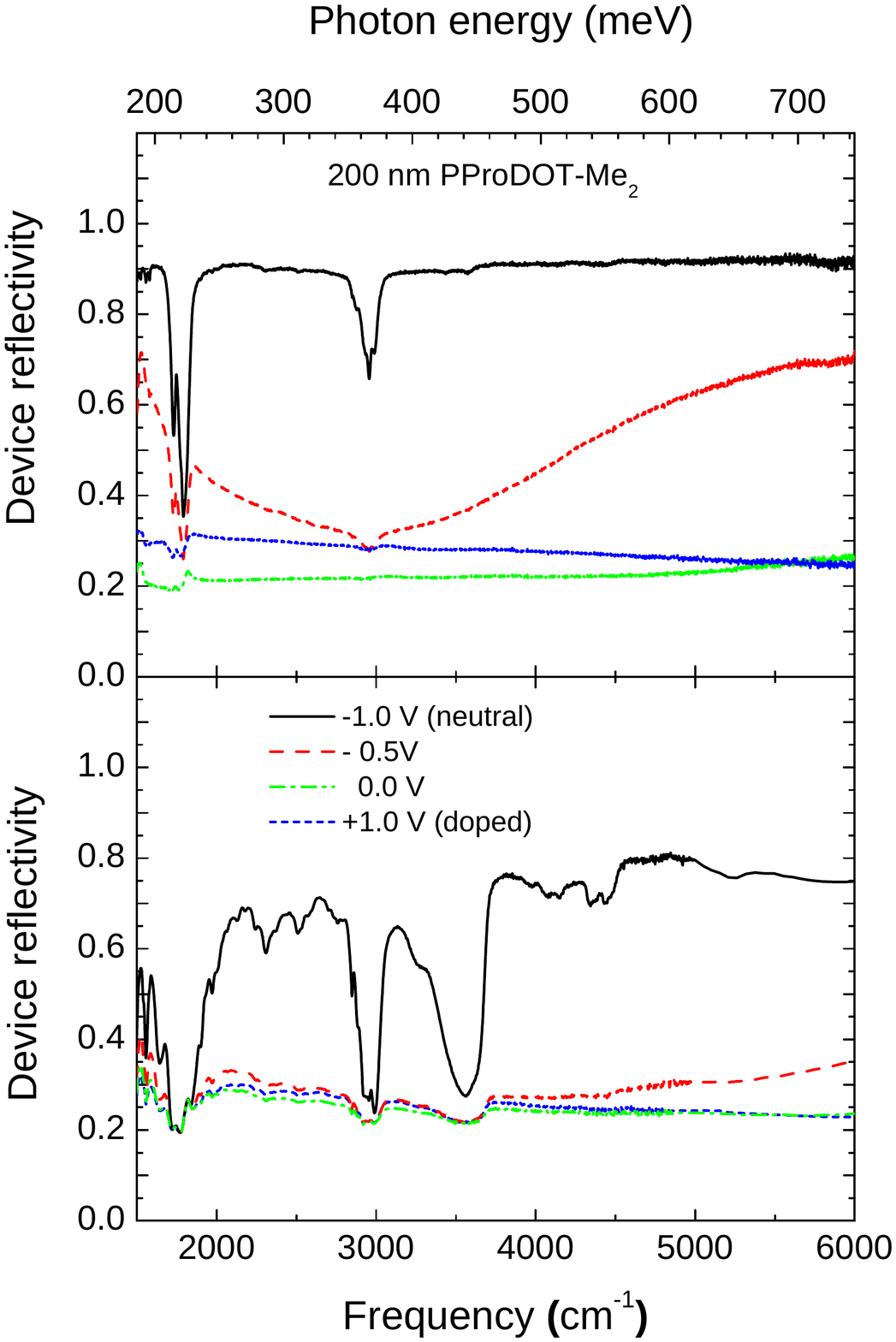}}%
 \vspace*{ -1.0cm}%
\caption{(Color online) Upper panel: Device reflectivity for a 200 nm thick PProDOT-Me$_2$ electrochromic cell with 125 $\mu$m electrolyte thickness at various potentials. Lower panel: Device reflectance of another electrochromic cell. These data are shown on a broader scale in Fig.~\ref{Pdmecc}. This electrochromic cell has water and much thicker electrolyte gel layer.} \label{NewEcc}
\end{figure}

\begin{figure}
\centering
 \vspace*{-0.5cm}%
 \centerline{\includegraphics[width=3.3 in]{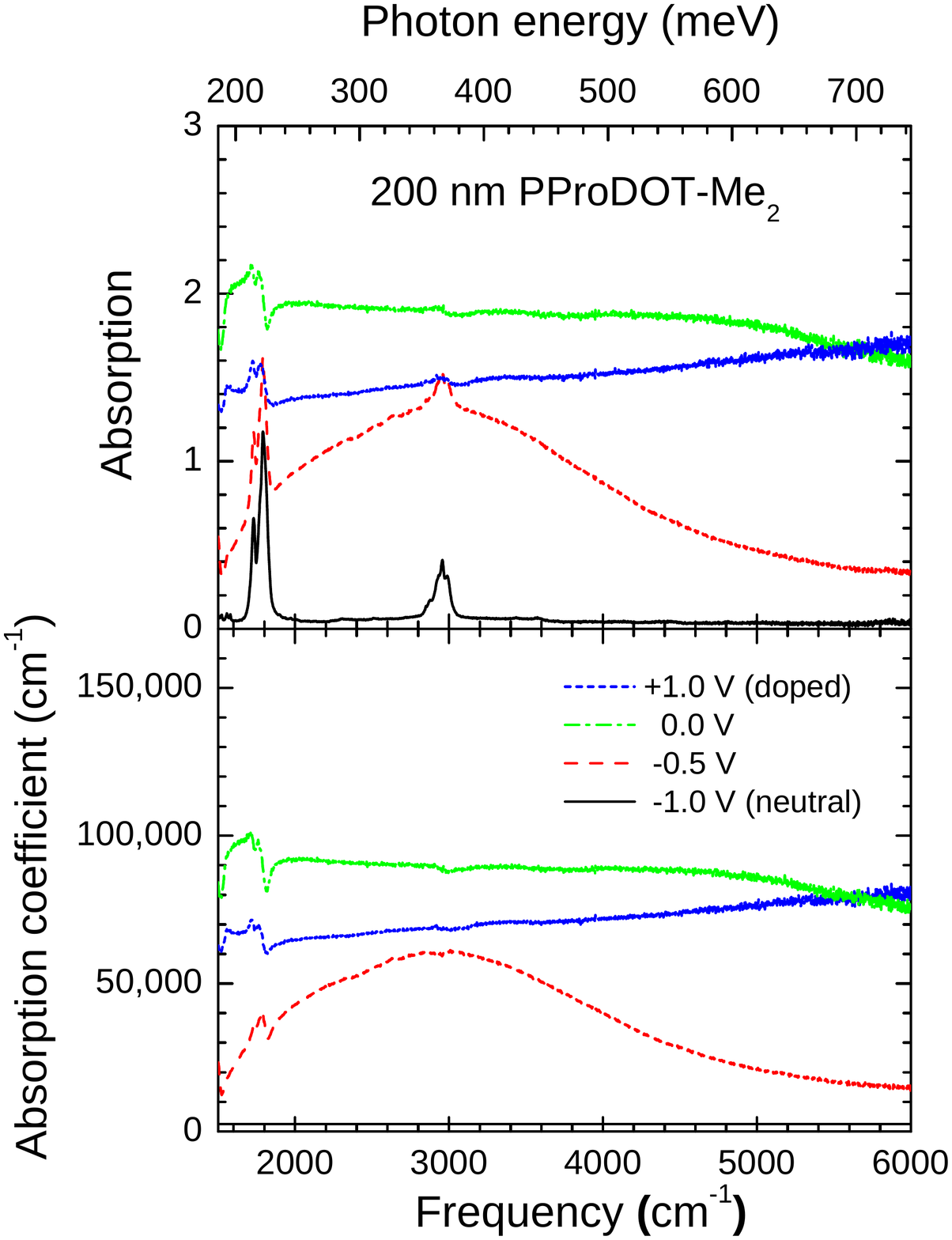}}%
 \vspace*{ -1.0cm}%
\caption{(Color online) The absorption (upper) and the absorption coefficient (lower) of 200 nm thick PProDOT-Me$_2$ at various potentials extracted using the procedure introduced in the previous section. Here, the absorption is $A(\omega)-(-\ln{R_{zs}(\omega)})$. We also multipled by a scale factor 1.9 to adjust the amplitude of the absorption coefficient.} \label{NewEccAnaly}
\end{figure}

The electrolyte gel transmittance is displayed in Fig.~\ref{EGel}. The gel was placed between two ZnSe windows separated by a 25 $\mu$m thicl Mylar spacer ring. After measuring the protected gel, the sample holder was opened to the air for one minute and the transmittance measured again. The absorbing O-H mode in the gel occurs at 3500 cm$^{-1}$ in the spectrum of the exposed electrolyte. This absorption is almost completely absent from the spectrum of the protected electrolyte, demonstrating that our process successfully prevented moisture from entering the gel. The spectrum also has two strong C-H modes, one near 1800 and the other near 2950 cm$^{-1}$. The stronger mode, near 1800 cm$^{-1}$, is saturated. The ratio of these two spectra was calculated and used to verify that the O-H mode contamination was due to water. The lower panel of Fig.~\ref{EGel} is a plot of the absorption coefficient $\alpha$ for the contaminant. Calculated from $\alpha_{water} \cong 1/d \ln(T_c/T_p)$  where $T_c$ and $T_p$ are the transmittance of the contaminated gel and protected gel respectively and $d$ is the thickness of the gel (25 $\mu$m). Water absorption bands can be seen clearly, the large peak near 3500 cm-1 (symmetric and antisymmetric stretch) and the smaller one at 1639 cm-1 (symmetric bend) are characteristic of a water absorption spectrum.\cite{banwell94}

The upper panel of Fig.~\ref{NewEcc} shows the reflectivity of an electrochromic cell made using the protected electrolyte gel as the potential across the sample varies from -1.0 V to +1.0 V. The lower panel shows device reflectivity spectra of the unprotected 200 nm PProDOT-Me$_2$ electrochromic cell at the same potentials. This new protected electrochromic cell showed much greater reflectivity changes in the 3500 cm$^{-1}$ range compared to electrochromic cells built with thicker, moisture-contaminated gel. That the spectra of our new cell have only small C-H absorptions is due to our ability to control and minimize the gel thickness. While we see clearly C-H absorption modes lowering reflectivity only near 1800 and 3000 cm$^{-1}$ for the new cell, we see C-H absorption modes at many regions for the contaminated cell. This water contamination occurred extremely quickly. After one full minute of exposure, the gel showed as much water absorption as a cell that had been constructed entirely in a moist environment.

We also applied the procedure introduced earlier in the paper to get the {\it in-situ} absorption and absorption coefficient of 200 nm PProDOT-Me$_2$ at various doping levels. The resulting absorption and absorption coefficient are shown in Fig.~\ref{NewEccAnaly}. The results agree well with the results in Fig.~\ref{EccAbsorp1b}. We can see the polaronic absorption near 3000 cm$^{-1}$ clearly even in the presence of the electrolyte gel contribution because the latter contribution is very weak. We note that there is no contribution from the ZnSe window in this spectral range.

Switching times for potential changes were extremely slow; on the order of one to two hours for a full switch. These long switching times are thought to be a consequence of thin gel layers and high pressures on the electrodes. This can be understood by noticing that the thin cuts in the upper electrode squeeze together when under pressure. A few simple tests were performed, showing that the device was capable of a full switch in 1.5 seconds when not under pressure. This would seem to suggest further study into the possibilities of quick switching in an airtight sample holder.

\section{Conclusions}

We studied electrochromism in dioxythiophene-based polymers, including {\it in-situ} device reflectivity, switching time, and lifetime (long-term redox switching stability). We investigated the doping-induced sub-bandgap features by controlling voltage between electrodes in the electrochemical cells. We find a growth of absorption due to in-gap polaronic states, which coexists with the $\pi$--$\pi*$ transition at intermediate doping levels. The lower polaronic band at lower frequency changes into the bipolaronic band while the polaronic mode at higher frequency disappears as doping increases.

The phase changes of the polymer films started from the slits cut into the upper layer. Because the lower polymer layer responds much more quickly than the upper polymer layer, the response or switching time of the cell mainly depends on the response time of the upper polymer film. This response time depends on the number of slits if other conditions are fixed (solvent, electrolyte, etc.). From the switching time experiment, we found that the response time of the p-doping (from neutral to p-doped phase) process is quicker (almost twice as fast) than the dedoping (from p-doped to neutral) process. This result suggests that Li$^{+}$ is the exchange ion for the p-doping and dedoping processes in the cell.

Another property determined was the long-term redox switching stability of the PProDOT-Me$_{2}$ electrochromic cell. The main factor for the long-term redox stability was the cell voltage, which must be lower than the overoxidation voltage of PProDOT-Me$_{2}$. Another factor was ensuring the gel electrolyte remained liquid. Due to the self-encapsulation by the PMMA in the gel electrolyte we could keep the cell liquid for a week in the laboratory environment. The result of the study showed that the PProDOT-Me$_{2}$ is very stable, because after 10,000 deep double potential switches the cell showed only about a 10\% decrease in the difference of device reflectivity by neutral and doped phases.

We also improved the performance of the electrochromic cell by removing water in the electrolyte gel. We were able to reduce the effect of the C-H absorption modes of the electrolyte gel by making the gel layer thinner, but we had to pay for this gain by accepting a very long cell switching time.

\begin{acknowledgments}

This work has been supported in part by the ARO through MURI grant
DAAD19-99-0316 and by the
and by the University of Florida Physics REU Site through NSF grants
DMR-9820518 and DMR-0139579.
J.H. acknowledges financial assistance from Pusan
National University Grant No. RIBS-PNU-2010-0075000.

\end{acknowledgments}

\end{document}